\documentclass[a4paper,11pt]{article}
\usepackage{jinstpub} % for details on the use of the package, please see the JINST-author-manual
\usepackage{lineno}
\title{\boldmath Development of a Central Trigger Processor board for the Advanced SiPM based camera of the CTA Large-Sized Telescopes}

\author[a]{A. Pérez-Aguilera}
\author[a]{M. Molina-Delicado}
\author[b]{T. Dietrich}
\author[a]{L.A. Tejedor}
\author[a]{J.A. Barrio}
\author[b]{A. Upegui}
\author[b]{Q. Berthet}
\author[a]{J. Buces}
\author[a]{D. Nieto}
\author[a]{D. Martín-Domínguez}
\author[c]{T. Miener}

\affiliation[a]{Instituto de Física de Partículas y del Cosmos, and EMFTEL Department, Universidad Complutense de Madrid (IPARCOS-UCM), E-28040 Madrid, Spain}
\affiliation[b]{inTECH, Hepia, University of Applied Sciences and Arts of Western Switzerland, 1202 Geneva, Switzerland}
\affiliation[c]{University of Geneva - Département de physique nucléaire et corpusculaire, 24 Quai Ernest Ansernet, 1211 Genève 4, Switzerland}

\emailAdd{alejpe15@ucm.es}

\abstract{We present ongoing work on the Central Trigger Processor board (CTPb), a trigger subsystem for the future advanced SiPM-based Large-Sized Telescope (LST) camera of the Cherenkov Telescope Array Observatory (CTAO). The camera will implement a fully digital trigger, exploiting the increased resolution to improve discrimination of low-energy $\gamma$-ray events from Night Sky Background noise. This approach aims to enhance telescope sensitivity while satisfying strict timing and data rate constraints. We describe the CTPb conceptual design and report initial results from hardware prototypes under evaluation within this next-generation trigger architecture.}


\begin{document}
\maketitle
\flushbottom

\section{Introduction}
\label{sec:intro}

The Cherenkov Telescope Array Observatory (CTAO) \cite{CTA} is an international collaboration aiming to build the next generation of Imaging Atmospheric Cherenkov Telescopes (IACTs). Its northern site is located at the Observatorio del Roque de los Muchachos on the island of La Palma, Spain. To cover the wide energy range of the $\gamma$-ray spectrum, different types of telescope are required. For the lowest energy range, four Large-Sized Telescopes (LSTs) with 23-m diameter reflectors are being constructed. The first of them, LST-1, has been successfully operating since 2019 and is already producing scientific results \cite{LST-CRAB}.

To detect brief Cherenkov flashes, lasting only a few nanoseconds, produced by the interaction of $\gamma$-rays with the atmosphere, the LST cameras are equipped with photomultiplier tubes (PMTs) \cite{PMT_module} and a very fast combined analog–digital readout and trigger system. The trigger systems are critical, as more than 99\% of the recorded data is useless, with its origins in light pollution, starlight, scattered moonlight, zodiacal light, airglow, etc. These noisy data, known as Night Sky Background (NSB), have a random nature and limit the sensitivity of the instrument. 

The current LST cameras are expected to last 15 years, while the CTAO operation is planned for 30 years. For this reason, the collaboration has taken the opportunity to design an advanced camera \cite{SiPM-ICRC}, based on silicon photomultipliers (SiPMs) and a fully digital readout and trigger system, capable of implementing more sophisticated trigger algorithms.

The camera will consist of 7987 SiPMs, arranged in a hexagonal geometry and grouped into clusters of 7 pixels each, for a total of 1141 clusters. Each Front-End Board (FEB) \cite{FEB} handles 49 SiPMs, performing three main functions: analog-to-digital conversion of SiPM signals, execution of the first trigger stage (L1), and transmission of event data to the DAQ system after a positive trigger. In total, the camera will require 163 FEBs. To fully capture the Cherenkov shower signals, the ADCs operate at a sampling rate of 1 GHz with 12-bit resolution. Without a trigger system to limit the events stored by the DAQ, the raw camera data rate would reach 95 Tb/s. The trigger architecture is completed by the Central Trigger Processor board (CTPb), which concentrates all trigger information from the camera and implements the next trigger stages. These next stages are the camera level trigger (L2) and the stereoscopic trigger level (L3).

The CTPb architecture is very close to converging to its final design, and the main technologies on which it is based are being tested.

\section{The Central Trigger Processor board}

The L1 algorithm consists of sums of the SiPM pixels digitized signals grouped in clusters of seven. Then, whenever the sum of the pixel signals within a cluster exceeds a predefined threshold, a local trigger is generated for that cluster. Therefore, the output product of the L1 trigger is an 1141 bit vector frame, 1 bit per cluster of 7 pixels, at a 1 GHz rate. As a result, each FEB will transmit 7 bits per nanosecond, which translates into a minimum 7 Gbps optical link. Taking into account possible overheads, 10 Gbps optical links are foreseen between the FEBs and the CTPb. AMD/Xilinx Kintex UltraScale Field Programmable Gate Arrays (FPGAs), with preference for the XCKU095 or XCKU115 models, were selected as the main platforms to receive incoming L1 data and process it, due to the density of Multi-Gigabit Transceivers (MGTs) and having enough logic resources to parallelize operations. To concentrate all optical links into the CTPb, Samtec Firefly optical transceivers, with 12 channels, were selected and tested in the first prototype, although other technologies more widely used and standardized, such as QSFPs, are not completely discarded. If finally used for the CTPb, a total of 15 of the Firefly connectors will be needed.

\begin{figure}[htbp]
\centering
\includegraphics[width=.4\textwidth]{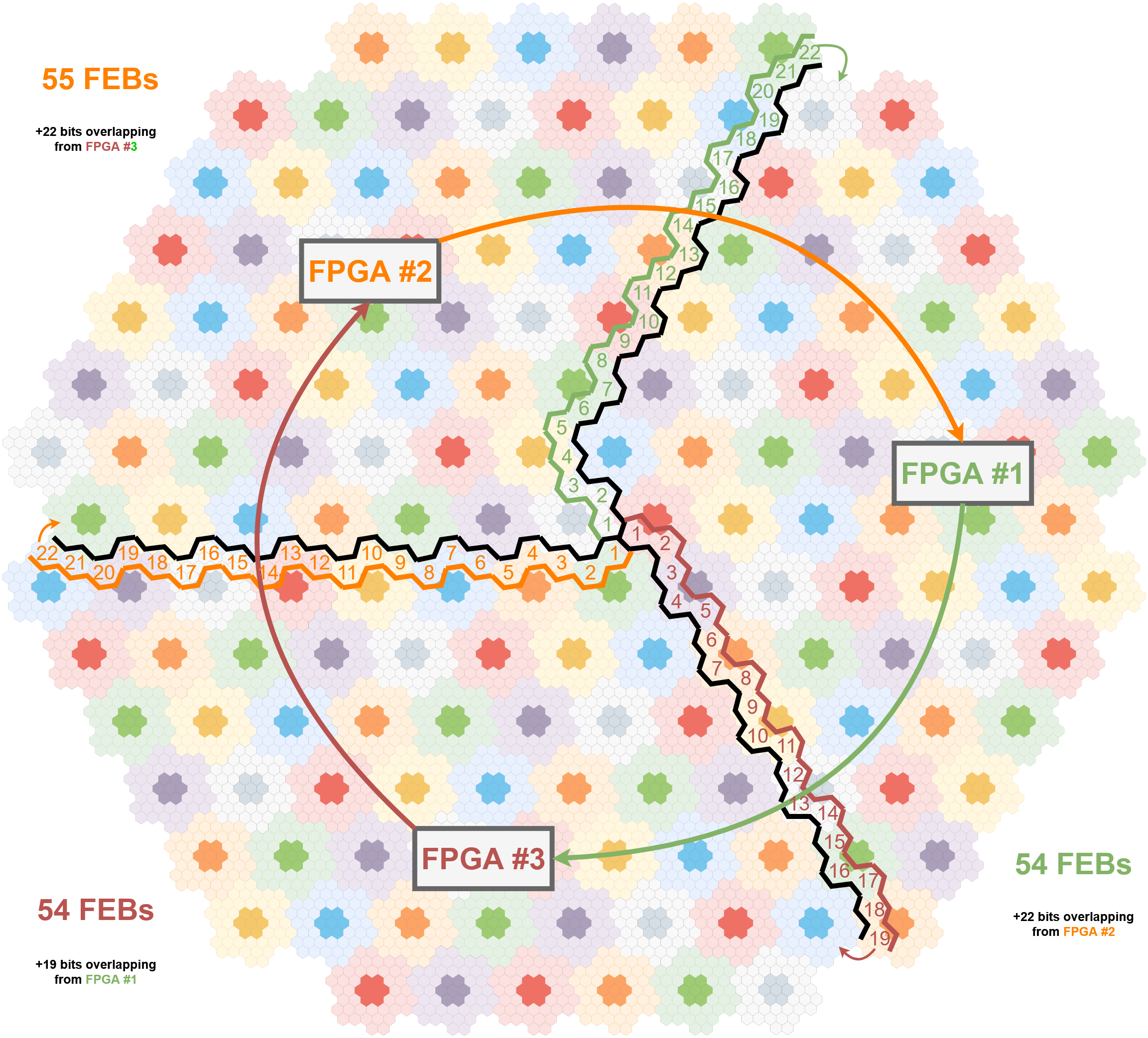}
\qquad
\includegraphics[width=.5\textwidth]{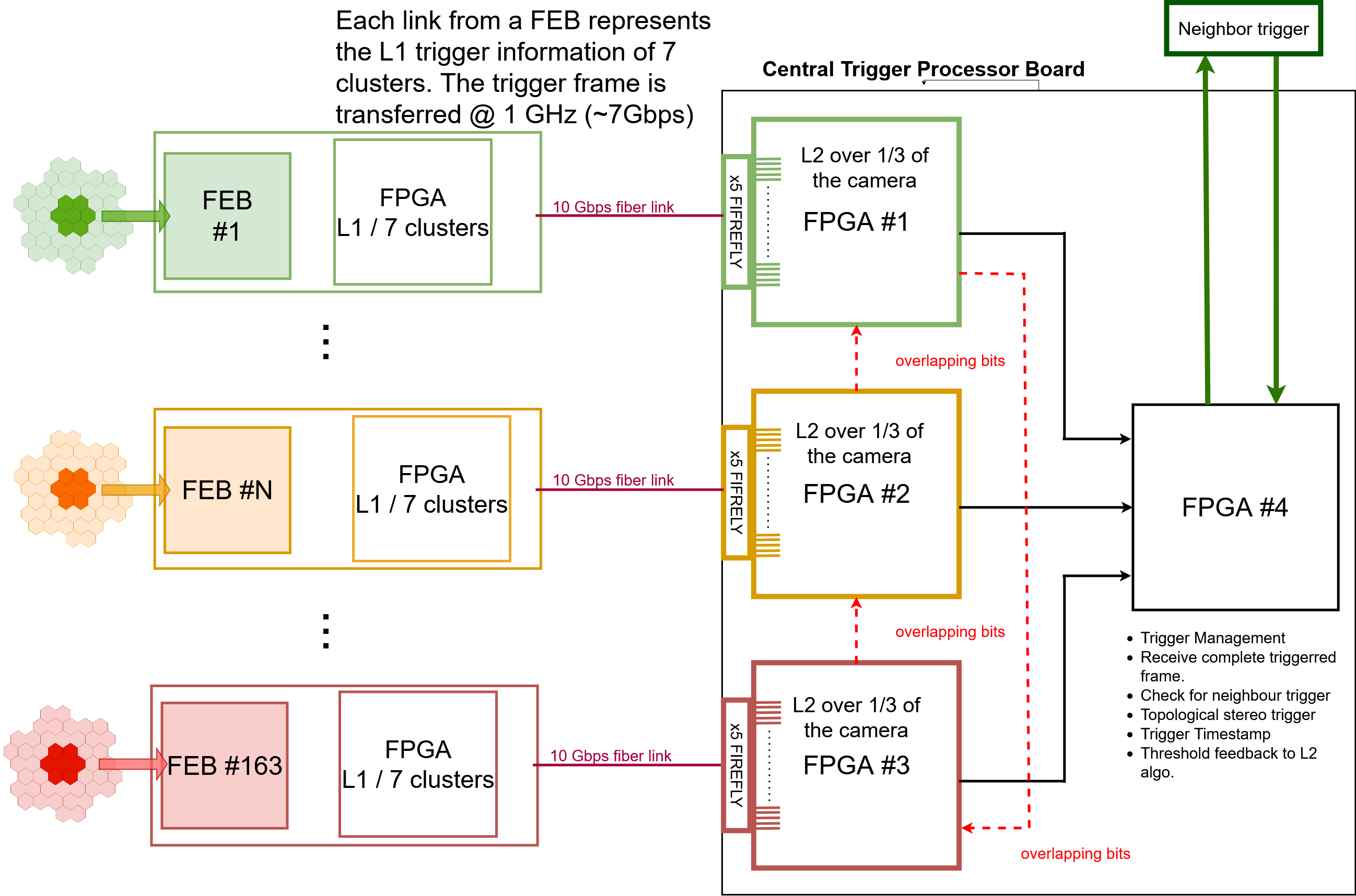}
\caption{(\textit{Left}) Pixels distribution in the camera, each colored area represents the corresponding pixels assigned to each FEB. The camera division shows the area corresponding to each processing FPGA at the CTPb. (\textit{Right}) FPGA disposition at the CTPb and the high-speed connections between them. \label{fig:CTP-ARCH}}
\end{figure}

As shown in Figure \ref{fig:CTP-ARCH}, the CTPb will house four FPGAs. The first three will receive the L1 data and run in parallel the L2 algorithm. CTPb will have to process an incoming data rate of 1141 bits per nanosecond, or 1141 Gbps. More than one FPGA is needed because of the limit of MGTs per chip. In this way, each FPGA will process a third of the camera L1 trigger information. An overlap of the regions is needed to ensure that events occurring in the boundary regions are not filtered. Two MGTs per FPGA are estimated to exchange the overlapping data between FPGAs. A fourth FPGA is needed to concentrate the results of the others and implement other tasks such as trigger time stamping, based on White Rabbit \cite{WR}, and the stereoscopic algorithm (L3), which will be based on the already proven concept of time and geometric coincidences between different telescope cameras \cite{TOPO}.

\section{L2 trigger algorithms}

Convolutional Neural Network (CNN) algorithms are being studied by the astroparticle physics community as one of the possible tools to analyze IACTs data. Due to their effectiveness in image processing and pattern recognition, CNNs have also been considered for the trigger system. A very small CTLearn \cite{CTLearn} model (a CNN with around 2000 parameters), able to process data cubes of 5 frames was translated to synthesizable firmware using the software package hls4ml \cite{hls4ml}. The best latency achieved, meeting timing constraints and resource usage, was 5.2 $\mu s$, which is equivalent to 1.1 Gbps, and not enough. Efforts are being made to improve the model architecture and its firmware implementation, by developing a custom C/C++ code implementation of the model with AMD Vitis platform. 

\begin{figure}[htbp]
\centering
\includegraphics[width=.6\textwidth]{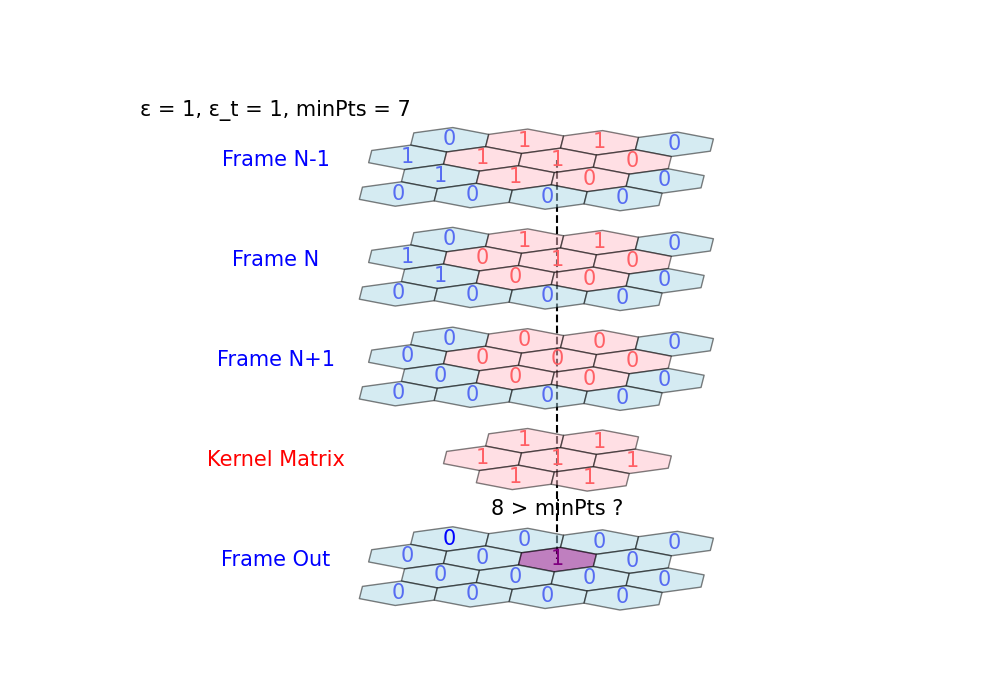}
\qquad
\includegraphics[width=.3\textwidth]{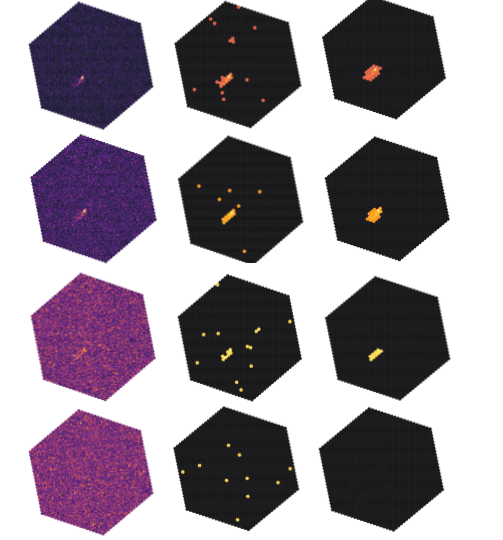}
\caption{(\textit{Left}) Hexagonal convolution over a L1 trigger region, taking into account three time frames. (\textit{Right}) Four consecutive frames of a simulated 68 GeV $\gamma$-ray event. The first column shows the raw digitized SiPM signals; the second shows the triggered clusters after the L1; the third shows the triggered clusters after the TDSCAN.\label{fig:L2_concept}}
\end{figure}

In parallel to CNNs, another algorithm has been tested with greater success: a modified version of DBSCAN (Density-Based Spatial Clustering of Applications with Noise) \cite{DBSCAN}. Given a set of nodes, DBSCAN groups together those that are geometrically close, which is very useful for detecting nearby triggered clusters in the camera. However, DBSCAN has some inconvenient disadvantages, the main one being its non-deterministic processing time per frame. The solution is the Trigger Distributed Spatial Convolution Accelerator Network (TDSCAN), which has a similar behavior to DBSCAN but with some conceptual differences: the need to only know if a cluster is present or not, fixed latency and a stream of images as input data. In practice, TDSCAN is basically a matrix convolution that checks if the number of neighbors of a certain point is above a threshold (Figure \ref{fig:L2_concept}). With a fully parallelized and pipelined design the algorithm can achieve a throughput of one frame per clock cycle. The algorithm has three configurable parameters: $\epsilon_{xy}$ is the size of the kernel (being $\epsilon_{xy} = 1$ a kernel of 7 neighbors), $\epsilon_t$ the number of frames before and after the computed frame (meaning that 5 frames are used if $\epsilon_{t} = 2$ ), and \emph{minPts} is the threshold above which the output bit is set to one, or trigger. Of course, greater $\epsilon_{xy}$ and $\epsilon_t$ mean more operations in parallel, which impacts resource usage in the FPGA.

\section{Setup and results}

The key components that determine the viability of the CTPb are the L2 trigger algorithm and the high-speed transmission links required to receive the L1 camera frames at a 1 GHz rate. If either of these elements failed or did not meet the system requirements, the entire trigger chain would need to be redesigned. To address this, two dedicated test platforms were developed.

\begin{figure}[htbp]
\centering
\includegraphics[width=.7\textwidth]{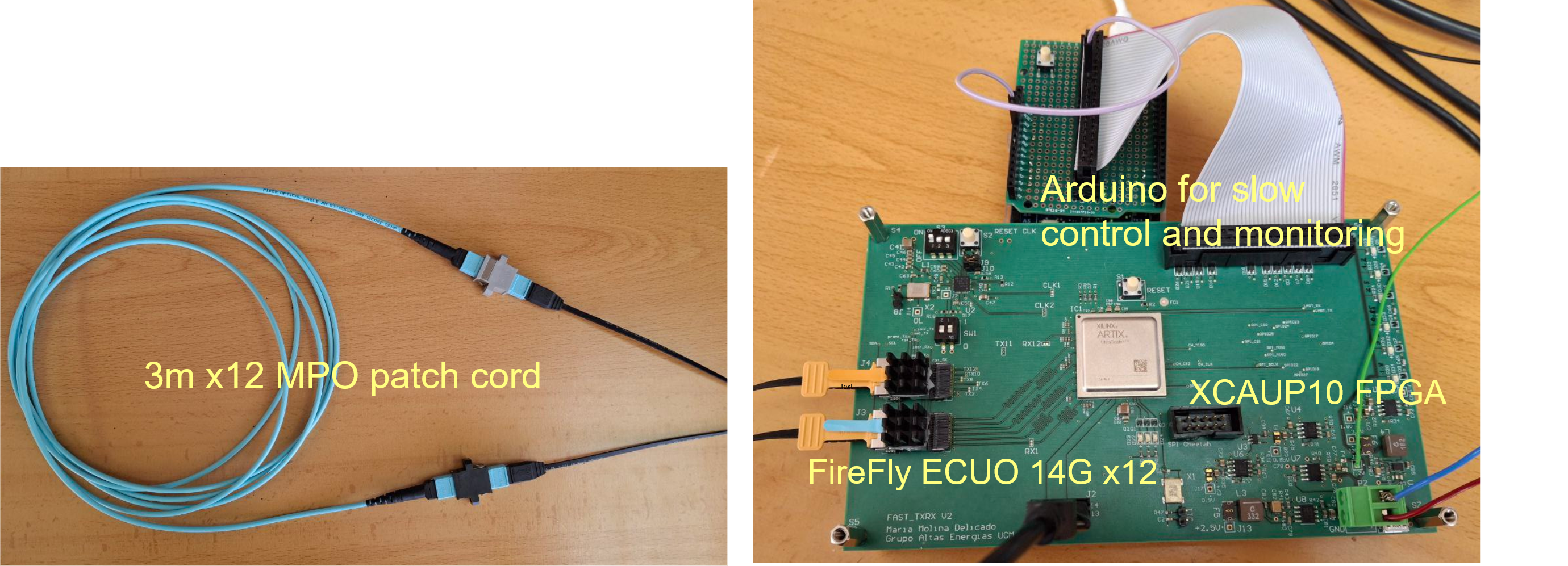}
\caption{Photograph showing the manufactured PCB with the cable to complete the data link loop for tests.\label{fig:PCB}}
\end{figure}

For the high-speed link tests, a custom PCB was manufactured (Figure~\ref{fig:PCB}). The main components of this board are two 12-channel Samtec FireFly transceivers, one hosting the optical receivers and the other the transmitters. An AMD/Xilinx Artix Ultrascale+ FPGA, cheaper than a Kintex Ultrascale but with the same type of MGTs, is included in the PCB, as its main tasks are generating and checking the transmitted data. In addition, a general-purpose connector was installed to provide monitoring and control of both the FPGA and the FireFly modules. The PCB substrate is ISOLA I-Tera MT-40, chosen for its dielectric constant and loss tangent values, suitable for frequencies up to 10 GHz. Although the PCB has relatively few components, the presence of the FPGA and multiple high-speed lines required a complex routing and layer stack-up. The board consists of ten layers, with some high-speed traces routed on the top layer and others through internal layers. 

Bit Error Rate (BER) tests were conducted to evaluate link performance. For these tests, a loop-back connection was made using an MTP/MPO fiber cable. Data was generated with a pseudo-random bit sequence (PRBS) module and checked upon reception. Tests were carried out at several data rates and with different channel configurations. A summary of the results is provided in Table~\ref{tab:ber_results}.

\begin{table}[htbp]
\centering
\caption{Summary of BER test results (no communication protocols applied).}
\begin{tabular}{|c|c|c|c|}
\hline
\textbf{Data rate (Gbps)} & \textbf{\# channels (Tx/Rx)} & \textbf{BER per channel} & \textbf{Power (W)} \\
\hline
6  & 4  & $<10^{-13}$ & 6.2 \\
\hline
6  & 12 & $<10^{-13}$ & 8.1 \\
\hline
10 & 4  & $<10^{-13}$ & 6.3 \\
\hline
10 & 12 & $<10^{-13}$ & 8.7 \\
\hline
\end{tabular}
\label{tab:ber_results}
\end{table}

\begin{figure}[htbp]
\centering
\includegraphics[width=1\textwidth]{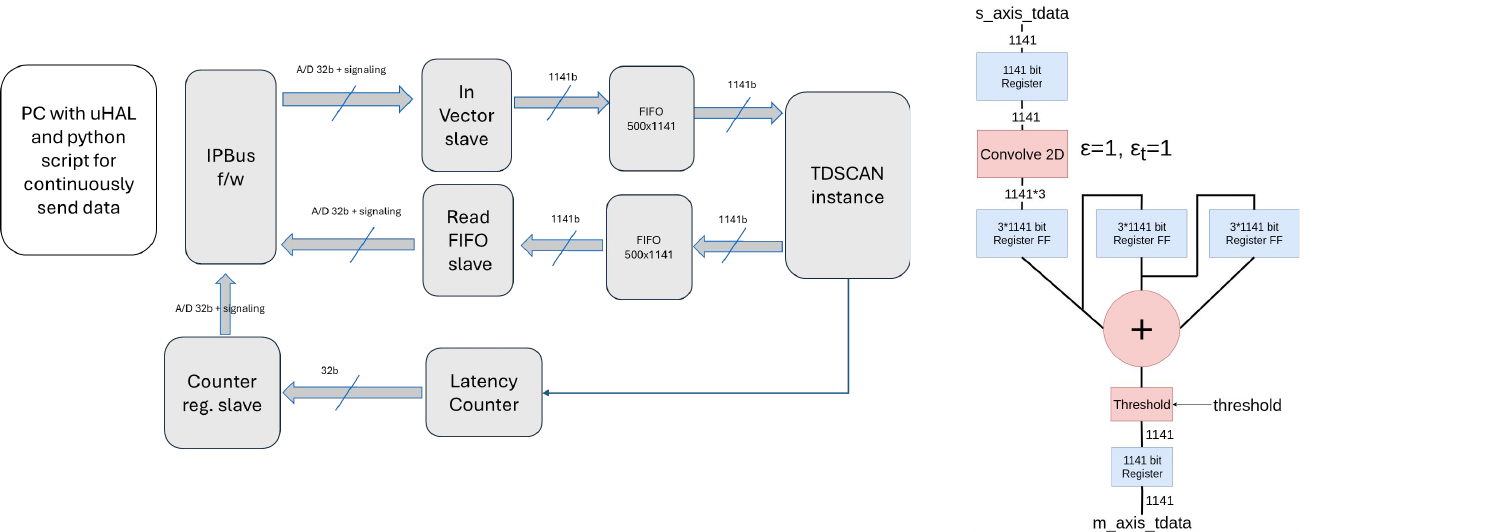}
\caption{(Left) Framework used to test a TDSCAN instance. Only the main modules are shown. Secondary modules, such as cross-clock domain synchronizers, are omitted for clarity. (\emph{Right}) Detail of the TDSCAN implementation, for these tests the values $\epsilon_{xy} = 1 $ and $\epsilon_t = 1$ were used, so the frame window is 3 and the convolution kernel is 7 pixels.\label{fig:tdscan-fw}}
\end{figure}

To validate the TDSCAN algorithm, a dedicated firmware was developed for the Alinx ACKU040 development board (Figure~\ref{fig:tdscan-fw}). The design relies on the IPBus protocol \cite{IPBus} to transmit simulated events to the FPGA and read the results. FIFO memories are used to buffer batches of data, while a counter measures the latency, and a Python script manages data transfer and result verification. The implementation leaves sufficient resources to accommodate multiple parallel TDSCAN instances, enabling operation at higher input data rates as well as integration of additional logic. The utilization summary per TDSCAN instance is: 22460 LUT (9.27\%), 65 LUTRAM (0.06\%), 22330 FF (4.61\%) and 53 BRAM (9.01\%). Timing closure was achieved with the TDSCAN and FIFO modules running at 400 MHz. A continuous stream of $10^{6}$ samples was processed with a total latency of 1,014,000 clock cycles, almost matching a processing time of one frame per clock cycle. Comparison against the same data processed with the same algorithm implemented in Python yielded a 1.4$\%$ discrepancy, which, together with the latency overhead, is attributed to the initialization of each 500-frame data batch. If the data stream were truly constant, without processing in batches, the discrepancies would disappear.

\section{Conclusion}

We have presented the development and validation of the key components for the Central Trigger Processor board (CTPb) for the advanced SiPM-based LST camera within the CTAO. The test benches demonstrate that the TDSCAN algorithm can be efficiently implemented in FPGA hardware, with moderate resource usage and latency fully compatible with the stringent requirements of the instrument. In addition, the high-speed optical links required by the system have been successfully validated up to 10 Gbps per channel with bit error rates below $10^{-13}$, confirming the robustness of the selected technologies. The modular architecture of the CTPb also paves the way for the integration of alternative approaches, such as machine-learning–based algorithms, which may further enhance trigger selectivity. 

\acknowledgments

This work is partially supported by the Spanish Ministry of Science and Innovation grants PID2022-138172NB-C42, PDC2023-145839-I00, and by the “Tecnologías avanzadas para la exploración del universo y sus componentes" (PR47/21 TAU) project funded by Comunidad de Madrid, by the Recovery, Transformation and Resilience Plan from the Spanish State, and by NextGenerationEU from the European Union through the Recovery and Resilience Facility

% Bibliography

%% [A] Recommended: using JHEP.bst file
%% \bibliographystyle{JHEP}
%% \bibliography{biblio.bib}

%% or
%% [B] Manual formatting (see below)
%% (i) We suggest to always provide author, title and journal data or doi:
%% in short all the informations that clearly identify a document.
%% (ii) please avoid comments such as "For a review'', "For some examples",
%% "and references therein" or move them in the text. In general, please leave only references in the bibliography and move all
%% accessory text in footnotes.
%% (iii) Also, please have only one work for each \bibitem.

\end{document}